\documentclass{ws-procs975x65-my} 

\usepackage{amssymb} 
\usepackage{amsfonts} 
\usepackage{amsmath} 
\usepackage{slashed} 

\newcommand{\beq}{\begin{equation}}
\newcommand{\eeq}{\end{equation}}
\newcommand{\beqa}{\begin{eqnarray}}
\newcommand{\eeqa}{\end{eqnarray}}
\newcommand{\nn}{\nonumber}
\newcommand{\half}{\frac{1}{2}}

\newcommand{\pbr}[2]{ \{ \hspace*{-2.6pt} [ #1 , #2\hspace*{1.4 pt} ] 
\hspace*{-2.6pt} \} }

\newcommand{\we}{\wedge}
\newcommand{\der}{\partial}

\newcommand{\ka}{\varkappa}
\newcommand{\hka}{\hbar\varkappa}

\newcommand{\Psib}{\overline{\Psi}}
\newcommand{\Phib}{\overline{\Phi}}

\newcommand{\what}[1]{\widehat{#1}}

\DeclareMathOperator{\Tr}{Tr} 

\newcommand{\rd}{\mathrm{d}} 
\newcommand{\ri}{\mathrm{i}} 

\newcommand{\omegab}{\bar{\omega}}
\newcommand{\gammab}{\bar{\gamma}}
\newcommand{\betab}{\bar{\beta}} 

\newcommand{\gammah}{\hat{\gamma\, }{\!}}

\begin{document} 

\unitlength = 1mm 

\title{Ehrenfest Theorem in Precanonical Quantization of Fields and Gravity}
\author{Igor V. Kanatchikov$^*$}

\address{School of Physics and Astronomy, University of St Andrews,\\
St Andrews KY16 9SS, Scotland\\
$^*$E-mail: ik25@st-andrews.ac.uk \\ 
}

\begin{abstract} 
\newcommand{\absab}{\tiny Precanonically quantized scalar field theory on curved space-time 
and the Einstein's pure gravity in Palatini vielbein formulation 
are compatible with the Ehrenfest theorem: 
the classical field equations in the De Donder-Weyl Hamiltonian form 
are obtained from the corresponding precanonical covariant 
Schr\"odinger equations as the equations of the expectation 
values of the corresponding operators. 
 } 
 
We discuss a generalization of the Ehrenfest theorem to the 
recently proposed precanonical quantization of vielbein gravity  
which proceeds from a  space-time symmetric 
generalization of the Hamiltonian formalism to field theory.  
Classical Einstein-Palatini equations are derived as 
equations of expectation values of precanonical quantum operators. 
The preceding consideration of an interacting scalar field theory on 
curved space-time 
 shows how the classical field equations 
 emerge 
 from the results of  precanonical quantization 
as the equations of expectation values of 
the corresponding quantum operators 
  It also 
allows us to identify  
the connection term in the covariant generalization of 
the precanonical Schr\"odinger equation with the 
spin connection.

\newcommand{\absbab}{\tiny 
Thus the  analog of the Ehrenfest Theorem is satisfied in precanonical quantization,  that confirms the 
consistency of the precanonical Schr\"odinger equation, the precanonical representation of operators, and the scalar product used for the 
calculation of expectation values with the dynamics in the classical limit.
} 
\end{abstract}

\bodymatter

\section{Introduction} 

In our previous papers\cite{my-quant,geom-quant} 
we have been exploring the potential of 
the De Donder-Weyl (DW) space-time symmetric generalization 
of the Hamiltonian formalism to field theory\cite{kastrup} 
as a basis of field quantization. 
The resulting {\em precanonical quantization} is based on the 
mathematical structures of the DW Hamiltonian formalism, 
such as the polysymplectic $(n$+$1)$-form (generalizing the symplectic 
$2$-form in mechanics to field theories in $n$ dimensions) 
and the Poisson-Gerstenhaber algebra of dynamical variables represented by a certain class of differential forms (generalizing the Poisson algebra of functions/functionals on the phase space in the canonical formalism of mechanics/field theory).\cite{my-pbr} 
It leads to the formulation of a quantum   theory of fields 
in terms of Clifford algebra-valued wave functions and 
operators on the finite dimensional space of field coordinates 
and space-time coordinates,\cite{my-quant,geom-quant} 
which generalizes the configuration space in mechanics 
to field theory without 
 the usual splitting to space and time. 
As the space-time coordinates enter the 
theory on the equal footing - as a multidimensional analog of the time variable - 
the corresponding space-time Clifford (Dirac) algebra generalizes 
the algebra of complex numbers in quantum mechanics (and its generalization to the continually infinite number of degrees of freedom, the standard QFT) 
to the present 
  description  of fields as ``multi-temporal" systems. 
The standard QFT in the functional Schr\"odinger representation\cite{hatfield} 
was shown to be 
derivable from precanonical quantization in the limit of the vanishing 
``elementary volume" $\frac{1}{\varkappa}$,\cite{my-schrod}  
which appears in the formalism 
of precanonical quantization when the dynamical variables represented by 
differential forms, which are infinitesimal volume elements, 
are represented by dimensionless elements of Clifford algebra. 

Here, 
 we first discuss the precanonical quantization of interacting scalar fields 
 on the curved space-time background and show how the classical field 
 equations in DW Hamiltonian form are derived from precanonical quantization as 
 the equations of expectation values of quantum operators. 
 Then we briefly outline the precanonical quantization of 
 vielbein general relativity in Palatini formulation and 
  sketch out the derivation of 
  the classical Einstein-Palatini equations in 
  DW Hamiltonian form from the 
  precanonical quantization of gravity as the equations 
 of expectation values of the corresponding operators. 
 That generalizes the Ehrenfest theorem  
 to the precanonical quantization of gravity.\cite{ehrenfest,review} 
 Further details of the derivation will be presented elsewhere.\cite{ehrenhest-gr}

 The concise presentation here is dictated by the limitations of the proceedings format of this paper. The reader is advised to consult 
 the previous papers by the author for the notation, terminology, 
 concepts, and the formalism underlying the precanonical quantization,  
 which are used here mostly without explanation. A companion paper\cite{mg14-lewand} may serve as a brief introduction.

\section{Ehrenfest Theorem in curved space-time} 

Let us  consider interacting scalar fields on a curved space-time background $g^{\mu\nu}(x)$:
\beq \label{scala}
{\mathfrak L} = \mbox{$\frac{1}{2}$} \sqrt{g}g^{\mu\nu} \der_\mu y^a \der_\nu y_a - \sqrt{g}\, V(y) , 
\eeq 
where $g\!:=\!|\det g_{\mu\nu}|$ and the parametric dependences from $x$ 
are henceforth not written down explicitly. 
Our purpose here is to present the DW Hamiltonian formulation, 
the resulting precanonical quantization of the system, and then to show 
that the latter reproduces the DW Hamiltonian equations 
as the equations on the expectation values of the corresponding operators. 
 
 The polymomenta and the DW Hamiltonian density obtained from (\ref{scala}):
\beq 
\mathfrak{p}^\mu_a 
 = \sqrt{g}g^{\mu\nu} \der_\mu y_a, 
\qquad 
{\mathfrak H} = \sqrt{g} H = \mbox{$\frac{1}{2\sqrt{g}}$} g_{\mu\nu} \mathfrak{p}_a^\mu \mathfrak{p}^{a\nu} + \sqrt{g} V(y) , 
\eeq
are densities  which 
parametrically depend on the space-time coordinates $x$ via $g^{\mu\nu}(x)$-s.   The DW Hamiltonian form of the Euler-Lagrange equations reads 
\newcommand{\ddd}{\mathbf{d}}
%
%
\beq \label{dw-curved}
\ddd_\mu \mathfrak{p}^\mu_a (x) = - \der_a \mathfrak{H}, 
\qquad 
\ddd_\mu y^a(x)  =  \der_{\mathfrak{p}^\mu_a} \mathfrak{H} , 
\eeq 
where $\ddd_\mu$ is the total differentiation w.r.t. $x^\mu$ 
and   
$\der_a := \der / \der y^a$, $\der_{\mathfrak{p}^\mu_a} 
:= \der / \der \mathfrak{p}{}^\mu_a$.

Precanonical quantization of this system leads to the representations 
\newcommand{\oldtextabb}{
The Poisson bracket operation defined by the weight $+1$ density valued polysymplectic structure (\ref{pscurv}) has a density weight $-1$, so that, for example,  
\beq
\pbr{\mathfrak{p}^\mu_a(x)}{y^b\varpi_\nu } = \delta^b_a\delta^\mu_\nu .
\eeq

The Dirac quantization rule  in curved space-time is also modified  
to make sure that density valued quantities are quantized as density 
valued operators of the same weight 
\beq
[\hat{A}, \hat{B}] = - \ri\hbar\sqrt{g} \what{\pbr{A}{B}} . 
\eeq
It leads to the following representations\cite{my-ehrenfest}  
} 
%
%
%
\beq \label{repsc}
\hat{\mathfrak{p}}{}^\mu_a = -\ri\hbar\ka\sqrt{g} \gamma^\mu \der_a , 
\qquad 
\hat{H} = - \mbox{$\frac{1}{2}$}\hbar^2\ka^2\der_a\der^a + V(y) , 
\eeq
where the curved-space $\gamma^\mu$-matrices are $x$-dependent:   
$\gamma^\mu\gamma^\mu+ \gamma^\mu\gamma^\mu = 2g^{\mu\nu}$,   
while the DW Hamiltonian operator turns out to be independent from $x$-s.  
The curved space-time covariant generalization of 
 the precanonical Schr\"odinger equation 
takes the form 
\beq \label{pse-curved}
\ri\hbar\ka\gamma^\mu \nabla_\mu\Psi (y,x) = \hat{H}\Psi (y,x) ,  
\eeq
where $\nabla_\mu \!:=\! \der_\mu + \omega_\mu  $ 
with the connection $\omega_\mu$ is a covariant derivative of 
Clifford algebra-valued wave functions.    
\renewcommand{\omegab}{\overline{\omega}}
 For the conjugate wave function 
 $\Psib \!:=\! \gammab{}^0\Psi^\dagger\gammab{}^0$ we obtain:
%
\beq \label{pse-conj}
\ri\hbar\ka \Psib ( \stackrel{\leftarrow}{\der_\mu} + \omegab_\mu ) \gamma^\mu 
= - \hat{H} \Psib , 
\eeq
where $\Psib := \gammab{}^0\Psi^\dagger\gammab{}^0$, 
$\gammab^I$ ($I=0,...,n-1$) are 
 flat-space Dirac matrices, such that 
$\gammab^I\gammab^J \!+\! \gammab^J \gammab^I \!=\! 2\eta^{IJ}$,  $\eta^{IJ}$ is 
the fiducial flat-space metric, 
$\omegab_\mu \!:=\!\gammab{}^0\omega_\mu^\dagger\gammab{}^0$,  
and a generalized Hermicity of $\hat{H}$ is assumed: 
 \mbox{$\hat{H}\! =\! \overline{\hat{H}}\!:=\!\gammab{}^0\hat{H}{}^\dagger\gammab{}^0$}.   

From (\ref{pse-curved}) and (\ref{pse-conj}) 
we obtain the covariant conservation law 
\beqa  \label{conserv-c}
\ddd_\mu \! \int\! \rd y\ 
\Tr\! \Big(\Psib \sqrt{g}\gamma^\mu\Psi \Big) 
=0 , 
\eeqa
provided 
$\omega_\mu$ satisfies 
the known property of the spin connection 
$\omega_\mu\!=\!\frac14 \omega_\mu^{IJ} \gammab_{IJ}$:\cite{bertlmann} 
\beq \label{theta-condition}
\der_\mu (\sqrt{g}\gamma^\mu) = \sqrt{g}\omegab_\mu\gamma^\mu + \sqrt{g}\gamma^\mu\omega_\mu . 
\eeq


Now, from (\ref{repsc}), (\ref{pse-curved}), (\ref{pse-conj}) and 
(\ref{theta-condition}), we obtain:   
\begin{align*} \label{curvav-p}
 \begin{split} 
\ddd_\mu \langle \hat{{\mathfrak p}}{}^\mu_a\rangle  = 
- \ri \hka\, \ddd_\mu\! \int\! \rd y\, \Tr\! \Big(\Psib \sqrt{g}\gamma^\mu \der_a \Psi \Big)
=&
- \ri\hka\int\! \rd y\, \Tr\! \Big(\der_\mu \Psib \sqrt{g}\gamma^\mu 
\der_a \Psi\\[-0pt] 
&\hspace*{-188pt}+\; \Psib \der_a\sqrt{g}\gamma^\mu \der_\mu  \Psi   
+  \Psib \der_\mu (\sqrt{g} \gamma^\mu ) \der_a \Psi  \Big) 
= - \int\! \rd y\, \Tr\! \Big( \Psib  [\der_a, \hat{\mathfrak{H}}]  \Psi\Big)  
= - \langle \der_a \hat{\mathfrak{H}}\rangle .
\end{split} 
\end{align*} 
Therefore,  the first DW Hamiltonian equation in 
(\ref{dw-curved}) is fulfilled on average.



Next, we note that (\ref{dw-curved}b) can be reformulated in terms 
of the covariant derivative and an $x$-dependent contravariant 
 $(n$--$1)$-volume element $\varpi^\mu:=g^{\mu\nu} \varpi_\nu$, 
 where
 $\varpi_\nu\!:=\! \imath_{\der_\nu} (dx^0\we...\we dx^{n-1})$:   
\beq \label{55}
\nabla_\mu (y^a\varpi^\mu) = \ddd_\mu (y^a \varpi^\mu) + 
\mbox{$\frac12$} y^a \der_\mu (\ln g) \varpi^\mu = 
\der_{\mathfrak{p}{^\nu_a}}\mathfrak{H} \varpi^\mu  . 
\eeq
Using the representation 
 $\what{\varpi}{}^\mu (x)=\frac{1}{\ka}\gamma^\mu (x)$, 
 and setting here $\hbar\!=\!1$, $\ka\!=\!1$, 
 we obtain 
\beqa
&&\ddd_\mu \langle \what{y^a \varpi^\mu} \rangle=
\int\! \rd y\ \Tr\! \Big( \der_\mu \Psib y^a\gamma^\mu \Psi + \Psib y^a\gamma^\mu \der_\mu\Psi +\Psib y^a (\der_\mu\gamma^\mu ) \Psi \Big) 
\nn \\[-0pt]
&&\!\!\!= 
\int\! \rd y\ \Tr\! \Big(\Psib\big(\ri\stackrel{\leftarrow}{\hat{H}}
-\omegab_\mu\gamma^\mu\big)  y^a \Psi 
- \Psib y^a \big(\ri\hat{H}+\ri\gamma^\mu\omega_\mu\big) \Psi 
+ \hka\Psib y^a (\ri\der_\mu\gamma^\mu)\Psi 
\Big) \nn \\[-0pt]
&&\!\!\!= \int\! \rd y\ \Tr\! \Big(\Psib [\hat{H}, iy^a] \Psi 
+\ri\Psib\big(\der_\mu\gamma^\mu-\omegab_\mu\gamma^\mu - \gamma^\mu\omega_\mu \big)y^a \Psi \Big) 
\nn \\[-0pt]
&&\!\!\! = \langle -i\der^a \rangle 
 - \mbox{$\half$}\langle y^a\gamma^\mu\rangle \der_\mu (\ln g) 
 = \langle \what{\der_{\mathfrak{p}{^\nu_a}}\mathfrak{H}\varpi^\nu} \rangle 
 - \mbox{$\half$}\langle y^a\gamma^\mu\rangle \der_\mu(\ln g) , 
\eeqa
 thus reproducing on average the second DW Hamiltonian equation in (\ref{dw-curved})  in the form (\ref{55}),  if 
  $\omega_\mu$ 
  satisfies the condition (\ref{theta-condition}) and hence can be identified with the spin connection. 

 Therefore, the analog of the Ehrenfest theorem 
for the precanonically quantized scalar fields on curved space-time 
is satisfied as the consequence of (i) the covariant precanonical Schr\"odinger 
equation (\ref{pse-curved}) and its conjugate (\ref{pse-conj}),   
(ii) the  precanonical representation of operators (\ref{repsc}), 
(iii) the definition of the scalar product related to the conservation law (\ref{conserv-c}), 
 %
 %
and (iv) the property (\ref{theta-condition}) of the connection 
term in (\ref{pse-curved}), which allows us to identify 
it with the spin connection  (the Fock-Ivanenko coefficients). 
 
\section{Precanonical quantization of 
 vielbein 
 gravity and the Ehrenfest Theorem} 

%
Here we follow our earlier work\cite{my-vielbein} 
(cf. Refs.~\refcite{my-metric} for an earlier work using the metric formulation).  
The Einstein-Palatini Lagrangian density with the cosmological term 
\beq \label{lagr}
{\mathfrak L}=  \mbox{$\ \mbox{$\frac{1}{\kappa}$}$} {\mathfrak e} e^{[\alpha}_I e^{\beta ]}_J 
\left(\der_\alpha \omega_\beta{}^{IJ} +\omega_\alpha {}^{IK}\omega_{\beta K}{}^J\right) + \mbox{$\mbox{$\frac{1}{\kappa}$}$}\Lambda {\mathfrak e} , 
\quad 
{\mathfrak e}:= (\det{||e^\mu_I}||)^{-1}, 
\eeq 
treats the vielbein components $e^\mu_I$ and the 
 spin connection coefficients $\omega_\alpha^{IJ}$ 
 as independent field variables. 
The DW Hamiltonian formulation leads to the 
polymomenta ${\mathfrak p}{}^\alpha_{e}$ 
and ${\mathfrak p}{}^\alpha_{\omega}$, 
and the DW Hamiltonian density $\mathfrak{H}\!=:\!\mathfrak{e}H$ derived 
from (\ref{lagr}):
\beq \label{constr}
\mbox{(a):}\;\; {\mathfrak p}{}^\alpha_{e_I^\beta}
\approx 0 , 
\;\;  \mathrm{} \;\;  
\mbox{(b):}\;\; {\mathfrak p}{}^\alpha_{\omega_\beta^{IJ}} 
\approx 
\mbox{$ \mbox{$\frac{1}{\kappa}$}$}
{\mathfrak e} e^{[\alpha}_Ie^{\beta ]}_{J },  
\quad\; 
\mbox{(c):}\;\; \mathfrak{H} =- \mbox{$\ \mbox{$\frac{1}{\kappa}$}$} {\mathfrak e} e^{[\alpha}_I e^{\beta ]}_J 
\omega_\alpha^{IK}\omega_{\beta K}^J 
- \mbox{$\frac{1}{\kappa}$}\Lambda{\mathfrak e}.   
\eeq 
The primary constraints (\ref{constr}a,b) are second-class, 
 because   
the brackets of their 
associated $(n$--$1)$-forms    
 \mbox{\small 
$
\mathfrak{C}_{e^\beta_I}
:={\mathfrak p}_{e^\beta_I}^\alpha\varpi_\alpha, 
\, 
\; 
\mathfrak{C}_{\omega_\beta^{IJ}}
:=  
\big( {\mathfrak p}{}^\alpha_{\omega_\beta^{IJ}} 
-
\mbox{$\ \mbox{$\frac{1}{\kappa}$}$}
{\mathfrak e} e^{[\alpha}_Ie^{\beta ]}_{J }\big) \varpi_\alpha  
$
}  
are not all vanishing: 
\beq \label{cbr}
 \hspace*{-1pt}\pbr{\mathfrak{C}_e}{\mathfrak{C}_{e'}} =0 
 = \pbr{\mathfrak{C}_\omega}{\mathfrak{C}_{\omega'}} ,\; 
 \;
 \pbr{\mathfrak{C}_{e^\gamma_K}}{\mathfrak{C}_{\omega_\beta^{IJ}}} 
 = - \mbox{\normalsize $\frac{1}{\kappa}$} \der_{e^\gamma_K} 
 \!\big( 
 {\mathfrak e} e^{[\alpha}_Ie^{\beta ]}_{J } 
 \big)  \varpi_\alpha  
=:  {\mathfrak{C}_{e^\gamma_K \omega_\beta^{IJ}}} .
\eeq 

%
%
%

\paragraph{Einstein's equations} 
are derived from (\ref{lagr}) by varying 
 $\omega$ and $e$ independently:  
\beq 
 \delta\omega\!: \nabla_\alpha\big(\mathfrak{e}e^\alpha_{[I} e^\beta_{J]} \big)= 0, 
 \quad 
 \delta e\!: 
 \der_{e^\mu_M}\big(
 {\mathfrak e} e^{[\alpha}_I e^{\beta ]}_J 
(\der_\alpha \omega_\beta{}^{IJ} +\omega_\alpha {}^{IK}\omega_{\beta K}{}^J ) 
+ 
\Lambda {\mathfrak e}\big) =0.  
\eeq
 The former equation defines the spin connection in terms of vielbeins, and the 
 latter one is the vacuum Einstein's equation in vielbein formulation. 
In the DW Hamiltonian formulation 
 these equations can be written in the form: 
 \beq \label{einst-dw}
 \mbox{(a):} \;\; \ddd_\alpha \mathfrak{p}^{\alpha}_{\omega_\beta^{IJ}} = 
 -\der_{\omega_\beta^{IJ}} \mathfrak{H} , \quad\; 
 \mbox{(b):} \;\; \mathfrak{C}^\alpha_{e^\gamma_K \omega_\beta^{IJ}} \ddd_{[\alpha} \omega^{IJ}_{\beta]} 
 =   
 - \der_{e^\gamma_K} \mathfrak{H} ,
 \eeq
 where $\mathfrak{C}_{e \omega} =: \mathfrak{C}^\alpha_{e  \omega} \varpi_\alpha$ 
 and the total (on-shell) derivative $\ddd$ also implies a restriction to the 
 subspace of constraints (\ref{constr}), 
 e.g.  $\ddd_\alpha \mathfrak{p}^{\alpha}_{\omega} = 
 \frac{\der \mathfrak{p}^{\alpha}_{\omega}(e)}{\der e}\frac{\der e}{\der x^\alpha}$. 
 Note that this formulation is inspired by the generalized Dirac analysis\cite{review,my-vielbein,my-dirac} 
 of the DW Hamiltonian system with the constraints (\ref{constr}).  
 
\paragraph{Quantization}
yields the representation of the operators of vielbeins 
and polymomenta:  
 $\hat{e}{}^\beta_I \!=\! -i \hbar\ka\kappa \bar{\gamma}^{J}\frac{\der}{\der \omega_{\beta}^{IJ}}$, 
$\; \hat{{\mathfrak p}}{}^{\alpha}_{\omega_\beta^{IJ}} 
 \!=\! - \hbar^2\ka^2\kappa 
\,\hat{\mathfrak e}\,
\bar{\gamma}^{KL}\frac{\der}{\der \omega_{[\alpha}^{KL}}
\frac{\der}{\der \omega_{\beta]}^{IJ}}$, 
which act 
  on 
   Clifford-valued  precanonical wave functions  
 $\Psi=\Psi(\omega^{IJ}_\alpha, x^\mu)$  
 on the (total space of the)  
 configuration bundle of spin connections over the 
 space-time. 
  For the operator of the DW Hamiltonian density 
$\mathfrak{H}\!=:\!{\mathfrak e}H$ restricted to the surface of 
constraints  $C$ given by  (\ref{constr}):  
${({\mathfrak e} H)|_C} \!=\! 
- {\mathfrak p}{}^{\alpha}_{\omega_\beta^{IJ}} \omega_\alpha^{IK} \omega_{\beta K}{}^{J} 
- \frac{1}{\kappa} \Lambda{\mathfrak e}$,  
we obtain (up to an ordering  \mbox{\footnotesize \raisebox{-2pt}{$\vdots$}}...\mbox{\footnotesize \raisebox{-2pt}{$\vdots$}})
\begin{equation} \label{hgrop}
\hat{H} = \hbar{}^2\ka^2\kappa\, \bar{\gamma}^{IJ} 
\mbox{\raisebox{-2pt}{$\vdots$}}\, 
\omega_{[\alpha}{}^{KM}\omega_{\beta] M}{}^L   
\der_ {\omega_{\alpha}^{IJ}}\der_ {\omega_{\beta}^{KL}}
\mbox{\raisebox{-2pt}{$\vdots$}}  
- \mbox{$\frac{1}{\kappa}$} \Lambda .  
\end{equation}

\paragraph{Precanonical Schr\"odinger equation} for quantum gravity:  
$i \hbar\ka \what{\slashed\nabla}  \Psi \!=\! \hat{H} \hspace*{-0.0em} \Psi$, 
where 
 $\what{\slashed\nabla} 
\!:=\! \widehat{\gamma}{}^\mu(\der_\mu+   
 \frac{1}{4} \omega_{\mu IJ} \bar{\gamma}^{IJ})$   
 and $\widehat{\gamma}{}^\mu \!:=\! \bar{\gamma}{}^I \hat{e}{}^\mu_I   
 \!=\! - i \hbar\ka\kappa \bar{\gamma}^{IJ} \der_ {\omega_{\mu}^{IJ}}$ 
 (cf. (\ref{pse-curved})), 
 now reads 
 \beq \label{wdw}
\bar{\gamma}{}^{IJ} \der_ {\omega_{\mu}^{IJ}}
 \der_\mu \Psi +  
 \bar{\gamma}{}^{IJ} 
\mbox{\raisebox{-2pt}{$\vdots$}}  
\big(  \mbox{$\frac{1}{4}$} \omega_{\mu KL}\bar{\gamma}^{KL} 
  - 
  \omega_{\mu}{}_{M}{}^{K}\omega_{\beta}^{ML} 
  \der_ {\omega_{\mu}^{KL}}
  \big) 
  \der_ {\omega_{\beta}^{IJ}} 
  \mbox{\raisebox{-2pt}{$\vdots$}}
   \Psi 
       + 
       \lambda \Psi = 0,     
 \eeq
where  $\lambda:= \frac{\Lambda}{(\hbar\kappa\ka)^2}$ is a dimensionless constant which combines three different scales: cosmological, Planck, and 
the 
 UV scale $\varkappa$ introduced by precanonical quantization.

 The conjugate wave function 
 $\Psib\!:=\!\gammab{\!}^0\Psi^\dagger\gammab{\!}^0$ obeys 
 \beq \label{wdwc}
 \der_ {\omega_{\mu}^{IJ}}
 \der_\mu \Psib \bar{\gamma}{}^{IJ} 
 - \Psib 
\mbox{\raisebox{-2pt}{$\vdots$}}  \big(  \mbox{$\frac{1}{4}$} \omega_{\mu KL}\bar{\gamma}^{KL} 
 \bar{\gamma}{}^{IJ}  
  + 
  \omega_{\mu}{}_{M}{}^{K}\omega_{\beta}^{ML} 
  \!\!\stackrel{\leftarrow}{\der}{\!}_ {\omega_{\beta}^{KL}}\big)
    \!\!\stackrel{\leftarrow}{\der}{\!}_ {\omega_{\mu}^{IJ}}
   \mbox{\raisebox{-2pt}{$\vdots$}}
       - \lambda \Psib = 0.     
 \eeq
 
 \renewcommand{\betab}{\bar{\gamma}{}^0}  

 \paragraph{The scalar product} 
 of precanonical wave functions is given by 
\beq \label{ovm}
\left\langle \Phi | \Psi \right\rangle 
:=  \Tr\! \int  \Phib \, \what{[\rd\omega]}_{} \Psi, \quad 
\what{[\rd\omega]} 
 =\ri^{\frac12 n(n+1)-1}
 \hat{{\mathfrak e}}{}^{- n(n-1)}\prod_{\mu, I<J} \rd \omega_\mu^{IJ},
\vspace{-5pt}\eeq 
where 
 \mbox{\small $\what{[\rd\omega]}$} is a Misner-like 
 diffeomorphism invariant 
 generalized-Hermitian 
 operator-valued measure on the 
fibers of the configuration bundle of spin connections over space-time 
 and 
 \beq
 \hspace*{-3pt}{\hat{\mathfrak{e}}{}^{-1}\!=\! 
 \frac{1}{n!} \epsilon^{I_1...I_n}\epsilon_{\mu_1...\mu_n} 
\hat{e
}{}^{\mu_1}_{I_1} ... \hat{e}{}^{\mu_n}_{I_n}}
=\frac{(-\ri)^n}{n!} \gammab{}^*  \epsilon^{I_1...I_n}\epsilon^{J_1...J_n}\epsilon_{\mu_1...\mu_n}
\der_{\omega_{\mu_1}^{I_1 J_1}}...\der_{\omega_{\mu_n}^{I_n J_n}} ,\hspace*{-3pt} 
 \eeq
where   ${\gammab}{}^* := \gammab{}^1\gammab{}^2...\gammab{}^n 
$.  
The numerical factor in (\ref{ovm}b) 
(which was omited in Refs.~\refcite{my-vielbein})   
comes from the generalized-Hermicity requirement: 
\mbox{\small $\what{{[\rd\omega]}}$}$\!=\!$
\mbox{\small $\overline{\what{[\rd\omega]}}$}.
 
The expectation values of operators are calculated as 
\beq
 \langle \hat{O}\rangle(x) = 
 \Tr\! \int \Psib (\omega,x) \mbox{\raisebox{-2pt}{$\vdots$}}\hat{O} \what{[\rd\omega]} 
 \mbox{\raisebox{-2pt}{$\vdots$}} \Psi (\omega,x) . 
\eeq


\paragraph{The conservation law}  
 derived from ({\ref{wdw}) and its conjugate  ({\ref{wdwc}) 
 (setting $\hbar\!=\!1\!=\!\ka$):   
\beqa \label{conserva}
\der_\mu \int \Psib  \hat{\mathfrak{e}} \hat{\gamma\, }{\!}^{\mu}[\what{\rd\omega}] \Psi 
&=&  \int \der_\mu \Psib  \hat{\mathfrak{e}} 
\hat{\gamma\, }{\!}^{\mu} [\what{\rd\omega}]\Psi + \int \Psib  \hat{\mathfrak{e}} \hat{\gamma\, }{\!}^{\mu} [\what{\rd\omega}]\der_\mu \Psi 
\nn \\[-0pt]
&&\hspace{-85pt}= \int 
\Psib \mbox{\raisebox{-2pt}{$\vdots$}}
\big(i\!\stackrel{\leftarrow}{\hat{H}} 
- \omega_\mu\! \stackrel{\leftarrow}{\hat{\gamma\, }{\!}^\mu}\!\big) 
\hat{\mathfrak{e}}[\what{\rd\omega}]\mbox{\raisebox{-2pt}{$\vdots$}} \Psi
- 
\int \Psib \mbox{\raisebox{-2pt}{$\vdots$}}
[\what{\rd\omega}] \hat{\mathfrak{e}} 
\big( i\hat{H} +\hat{\gamma\, }{\!}^\mu\omega_\mu \big)\mbox{\raisebox{-2pt}{$\vdots$}} \Psi  
  \\[-0pt]
 &&\hspace{-11pt}= ... = 
 \langle \mbox{\raisebox{-2pt}{$\vdots$}}\hat{\mathfrak{e}}\,\omegab_\mu\hat{\gamma\, }{\!}^\mu + \hat{\mathfrak{e}}\hat{\gamma\, }{\!}^\mu\omega_\mu \mbox{\raisebox{-2pt}{$\vdots$}} \rangle , \nn 
 \eeqa
is  equivalent to the fulfillment on average 
of the property (\ref{theta-condition}) 
of curved-space Dirac matrices and spin connection,    
because the left hand side of (\ref{conserva}) 
is $\der_\mu\langle \hat{\mathfrak{e}} \hat{\gamma\, }{\!}^{\mu} \rangle$. 
As eq.~(\ref{theta-condition})  is a consequence of the first 
of the Einstein-Palatini equations in (\ref{einst-dw}),  
the result in (\ref{conserva}) can be seen as 
a first indication that our precanonical 
quantization  is consistent in the classical limit 
at least with the classical geometry underlying GR. 

Note, however, that 
we had to omit several essential intermediate details 
in the calculation of (\ref{conserva}). 
 The most important one is that 
the terms with $\Lambda$ do not cancel on their own, 
that would lead to the violation of the 
covariant probability conservation law due to the cosmological constant. 
Though it might sound plausible, we think that the true message here 
is that the cosmological constant, which was introduced in the classical 
Lagrangian (\ref{lagr}),  
should be cancelled by the constants generated by a proper choice of 
the ordering inside $\mbox{\small \raisebox{-2pt}{$\vdots$}} ... \mbox{\small \raisebox{-2pt}{$\vdots$}}$. This cancellation yields 
a prediction of the admissible value of $\Lambda$, 
though  in terms of an yet unspecified scale 
$\varkappa$ introduced by precanonical quantization,\cite{my-vielbein} 
and fixes the ordering ambiguity in $\hat{H}$ by requiring 
 $\mbox{\small \raisebox{-2pt}{$\vdots$}} \omega\omega\der_\omega\der_\omega \mbox{\small \raisebox{-2pt}{$\vdots$}}$ to be anti-Hermitean.



\paragraph{Ehrenfest theorem for the Einstein's equations} can be 
obtained now for the DW Hamiltonian form of the latter (\ref{einst-dw}). 
By proceeding similarly to (\ref{conserva}), we~obtain 
\beqa
\ddd_\alpha\langle \hat{\mathfrak{p}}^\alpha_{\omega_\beta^{IJ}}\rangle 
&=&  -\ri\hbar\ka \der_\alpha \Tr\! \int \Psib 
\hat{\mathfrak{e}}\hat{\gamma\, }{\!}^\alpha\der_{\omega_\beta^{IJ}} 
\what{[d\omega]}\Psi = ...
 = - \langle \der_{\omega_\beta^{IJ}} {\what{\mathfrak H}} \rangle , 
\eeqa
that reproduces the first of the Einstein-Palatini equations, 
eq.~(\ref{einst-dw}a),  
on average. 

The Einstein's equations proper, eq.~(\ref{einst-dw}b),  
are more tricky to obtain as the Ehrenfest-type statement. 
 Using  
$\mathfrak{C}^\alpha_{e^\gamma_K \omega_\beta^{IJ}} \ddd_{[\alpha} \omega^{IJ}_{\beta]} 
 \!=\!   \ddd_{\alpha} \big(\mathfrak{C}^\alpha_{e^\gamma_K \omega_\beta^{IJ}}\omega^{IJ}_{\beta}\big) \!-\! 
 \ddd_{\alpha}\big(\mathfrak{C}^\alpha_{e^\gamma_K \omega_\beta^{IJ}}\big)  \omega^{IJ}_{\beta}$,   
   eq.~(\ref{einst-dw}a), the expression of $\mathfrak{C}_{e\omega}$ in 
   (\ref{cbr}),  and the constraint (\ref{constr}b),  
the Einstein's equations can be cast in the form more suitable for us:  
\beq \label{einstu}
  \ddd_{\alpha} \big(\mathfrak{C}^\alpha_{e^\gamma_K \omega_\beta^{IJ}}\omega^{IJ}_{\beta}\big) 
  =  \der_{e^\gamma_K} \mathfrak{H}. 
\eeq

The construction of the operators involved in (\ref{einstu}) 
makes use of the observations that 
(i) 
 $\what{\der_{e^\gamma_K} A} \!=\! q [\ri\, \omega_\gamma^{KL}\gammab_L,\hat{A}]$,  
where $q$ is a numerical/sign factor, 
(ii) $\mathfrak{C}^\alpha_{e^\gamma_K\omega_\beta^{IJ}}
\omega^{IJ}_{\beta}\!\approx\! - \der_{e^\gamma_K} \big( \omega^{IJ}_{\beta}\mathfrak{p}^\alpha_{\omega_\beta^{IJ}} (e) \big)$, 
and (iii) for any $A^\alpha$, $\ddd_\alpha A^\alpha \!=\! \ddd_\alpha(\rd x^\alpha\bullet A^\nu\varpi_\nu)$.\footnote{$A\bullet B:=*^{-1}(*A\we *B )$ is the product 
operation in the Poisson-Gerstenhaber algebra of brackets which 
underlies the precanonical quantization.\cite{}}  
Using the representations 
$\what{dx^\nu\bullet}= (-1)^{n-1}\ka\gammah^\mu${} 
(cf. Ref.~\refcite{ehrenfest})  
and $\what{\mathfrak{p}_\omega{}^\nu\varpi_\nu} = -\ri\hbar \der_\omega$,   
we get 
$\what{\mathfrak{C}^\alpha_{e^\gamma_K\omega_\beta^{IJ}}\omega_\beta^{IJ}} 
 =   
- \ri\hbar q \mbox{\raisebox{-2pt}{$\vdots$}}\gammah^\alpha  \omega_\gamma^{KL}\gammab_L 
\mbox{\raisebox{-2pt}{$\vdots$}} $.  

Now, using  the precanonical Schr\"odinger equation (\ref{wdw}) 
and its conjugate  (\ref{wdwc}), we obtain  
\beqa 
\ddd_{\alpha} \what{\langle\mathfrak{C}^\alpha_{e^\gamma_K \omega_\beta^{IJ}}\omega^{IJ}_{\beta}\rangle} 
&=& \ddd_\alpha\Tr\!\int \Psib  \gammah^\alpha 
q(-\ri\hbar)\mbox{\raisebox{-2pt}{$\vdots$}} \omega_\gamma^{KL}\gammab_L \what{[d\omega]}\mbox{\raisebox{-2pt}{$\vdots$}}\Psi
 = ... =  \langle \mbox{\raisebox{-2pt}{$\vdots$}}  \what{\der_{e^\gamma_K}\mathfrak{H}}\mbox{\raisebox{-2pt}{$\vdots$}}\rangle ,
\eeqa 
thus reproducing the Einstein's equations in the form (\ref{einstu}) 
as the equation of expectation values of precanonical quantum operators. 


\section{Conclusion} 

The ability of precanonical quantization to reproduce correctly the classical
field equations as the equations of  expectation values of quantum operators, 
i.e. the validity of the Ehrenfest theorem,  
can be considered as a consistency test of the 
(i) precanonical representation of operators, 
(ii) precanonical analog of the Schr\"odinger equation and 
(iii) the prescription of the calculation of expectation values of 
operators using the Clifford algebra-valued precanonical wave functions. 
In the case of scalar fields on curved background the generalization 
of the  Ehrenfest theorem identifies the connection term in the Dirac operator 
of the precanonical  Schr\"odinger equation as the spin connection. 
This observation allows us to proceed with the precanonical 
quantization of general relativity with more confidence. 

The sketch of the derivation of the Einstein-Palatini equations for 
vielbein gravity from the quantum theory of gravity obtained 
by precanonical quantization of vielbein general relativity indicates, 
in particular,  that the value of the cosmological constant 
is fixed by a suitable ordering of operators and the consistency with 
the local covariant probability conservation, which also coincides  
with the known property of the spin connection and curved 
Dirac matrices fulfilled on average.  
A very rough estimation\cite{my-vielbein}{}$^c$    
of the value of the dimensionless parameter $\lambda=\mbox{$\frac{\Lambda}{(\hbar\kappa\varkappa)^2}$}\sim n^3$, 
which originates from the 
re-ordering of operators $\omega$ and $\der_\omega$ in precanonical Schr\"odinger equation, has lead us to an unexpected 
conclusion that the parameter $\varkappa$ of pre-canonical 
quantization, which is  consistent with the observed values 
of $\Lambda$, $G$ and $\hbar$, is at the nuclear scale.  
However, this preliminary consideration of the pure gravitational 
contribution to $\Lambda$ neglects the matter fields.  
We need an independent argument regarding the estimation 
of the scale of $\varkappa$, if it is a universal scale, 
which might be provided e.g. by a precanonical 
quantization approach to the mass gap problem in quantum Yang-Mills theory 
(see Ref.~\refcite{ym} for a work in this direction).

\section*{Acknowledgments} 
I would like to express my gratitude to the School of Physics and Astronomy 
of the University of St Andrews for its kind hospitality and the possibility 
to use its facilities at any time of the day or night, which made the progress 
reported here possible.  


 \end{document}